\documentclass[preprint,showpacs,preprintnumbers,amsmath,amssymb]{revtex4}
\usepackage{graphicx}
\usepackage{dcolumn}
\usepackage{bm}

\begin{document}


\title{Method of measurements with random perturbation:\\Application in photoemission experiments}

\author{D. S. Fedin and O. N. Granichin}
\affiliation{Department of Mathematics and Mechanics, St. Petersburg State University, 198904 St. Petersburg, Russia}

\author{Yu. S. Dedkov\footnote{Corresponding author. Electronic address: dedkov@physik.phy.tu-dresden.de} and S. L. Molodtsov}
\affiliation{Institut f\"ur Festk\"orperphysik, Technische Universit\"at Dresden, 01062 Dresden, Germany}

\date{\today}

\begin{abstract}

We report an application of a simultaneous perturbation stochastic approximation (SPSA) algorithm to filtering systematic noise (SN) with non-zero mean value in photoemission data. In our analysis we have used a series of 50 single-scan photoemission spectra of W(110) surface where randomly chosen SN was added. It was found that the SPSA-evaluated spectrum is in good agreement with the spectrum measured without SN. On the basis of our results a wide application of SPSA for evaluation of experimental data is anticipated.

\end{abstract}


\maketitle

Very often measurements of physical quantities are considerably hindered by observation of noises of unknown nature. This problem appears in many industrial as well as scientific applications, such as pattern recognition, product quality improvement, control of heavy ion beams, probing of electronic structure of solids and molecules by spectroscopic techniques, \textit{etc}. Traditionally, experimental noises of different origins are assumed to be independent from each other and characterized by zero-mean values. These assumptions are frequently hard to justify. But without them, the validity of many algorithms is questionable in applications. For example, it is known that the standard ``least-squares method" or the ``maximum likelihood method"~\cite{Brandt:1983} give wrong estimates if the observed noise has an ``unknown-but-bounded" deterministic nature or it is a ``dependent" sequence  from probabilistic point of view. 

Let us consider photoemission (PE) measurement~\cite{Kevan:1992,Hufner:2003} where systematic noise (SN) of unknown origin appears at some particular kinetic energies of photoelectrons. Such noise may introduce additional spectral features that can not be expected from electronic structure of sample. If this noise is not zero-mean it cannot be eliminated by simple increase of number of scans in PE experiment. One of the effective ways to deal with such noise is using of new mathematic algorithms.

Mathematical algorithms for search and optimization play a severe role in finding best options to solve many problems not only in physics, but in engineering, business, medicine, as well as in other natural and social sciences. In the case of prior information ambiguity, recursive algorithms are most effective among many other approaches. Such algorithms start with an initial ``guess" of a solution, and this assessment is updated on an iteration basis with the purpose of improving the measured (observable) objective function of sample. In most practical problems, the relevant solution depends on more than one quantity, leading to multivariate optimization problem of minimizing or maximizing the objective function dependent on multiple variables. There is much interest in recursive optimization algorithms that rely on direct measurements of the objective function only, not direct probing of its gradient. Such algorithms have the advantage of not requiring detailed modeling information describing links between parameters to be optimized and the objective function. Many systems involving human beings or computer simulations that are difficult to treat analytically, could potentially benefit from applications of this kind of optimization approaches.

One particular optimization algorithm that has attracted considerable international attention in the recent past is the simultaneous perturbation stochastic approximation (SPSA) method (see for review~\cite{Spall:1998,Granichin:2003}). Likewise simulated annealing or genetic algorithms, SPSA uses only objective function measurements, that is of decisive importance in experiments where direct probing of the gradient of the objective function is rather difficult or not possible at all. In difference to other methods, SPSA is especially efficient to high-dimensional problems in terms of providing a good solution for a relatively small number of measurements of the objective function. Recently this algorithm was successfully applied to a number of tasks like queuing systems~\cite{Fu:1997} or control of heavy ion beams~\cite{Hopkins:1997}.


In the present study we demonstrate for the first time application of a SPSA algorithm for analysis of PE spectra, which include a noise with non-zero mean value. PE spectra of W(110) surface were collected upon excitation with two \textit{independent} photon sources, one of which was used as a probe and second one as a source of the noise. It was shown that the spectrum obtained after application of the SPSA algorithm to a series of fifty PE single-scan spectra is in good agreement with the spectrum measured without SN. On the basis of these results we conclude that exploiting of SPSA can be useful for analysis of any experimental spectroscopic data. We can expect a wide application of this method for filtering systematic noises that can appear in many kinds of measurements or experiments.

Photoemission spectra~\cite{Kevan:1992,Hufner:2003} were taken from a W(110) single crystal kept at room temperature. Experiments were performed in the setup based on the hemispherical energy analyzer (SPECS PHOIBOS 150)~\cite{specs}. The overall-system energy resolution accounting for the thermal broadening was set to 150\,meV and electrons were collected in angle-integrated mode around the surface normal. The base pressure was in the range of $1\times10^{-10}$\,mbar. Prior to experiment, the W(110) crystal was carefully cleaned by repeated cycles of heating up to $1300^\circ$\,C in oxygen ambient pressure of $5\times10^{-8}$\,mbar for 15\,min each and subsequent flashing up to $2300^\circ$\,C. After such procedure the crystal was kept in vacuum for 24 hours in order to passivate the surface of the crystal via absorption of residual gases in the experimental chamber. This step was necessary in order to stabilize the crystal surface in the long-term surface-sensitive PE experiment. As an excitation light sources we have used He\,II$\alpha$ resonance line ($h\nu=40.8$\,eV) and Al\,$K\alpha$ emission line ($h\nu=1486.6$\,eV) in order to generate studied PE signal (objective function) and independent non-zero mean noise, respectively. The scheme of the experiment is shown in Fig. 1(a). PE spectra were collected in the range of $25.8-37.8$\,eV kinetic energies of emitted photoelectrons. In this case, the He\,II$\alpha$ radiation of fixed intensity was used to produce each time single-scan PE spectrum of the valence band of W(110) surface (Fig. 2, open circles). Thereby different single-scan spectra were excited with different He\,II$\alpha$ radiation intensities which were randomly selected. The photocurrent emitted in this process can be expressed as
\begin{equation}
j(E_{kin})=I\cdot DOS(E_{kin})\cdot {d\sigma(E_{kin})}/{d\Omega},
\end{equation}
where $I$ is the intensity of the light source (He\,II$\alpha$), $DOS$ denotes the electronic density of states of the W(110) surface, ${d\sigma(E_{kin})}/{d\Omega}$ is the cross-section of the photoemission process, $E_{kin}=h\nu-W-E_B$ stands for the kinetic energy of the photoelectron ($W-$work function of the material, $E_B-$binding energy of the electron in the solid). Since ${d\sigma(E_{kin})}/{d\Omega}$ is practically constant in the small energy range, the total photocurrent can be written as
\begin{equation}
j(E_{kin})=I\cdot const \cdot DOS(E_{kin}).
\end{equation}
It is proportional to density of states and to intensity of incoming radiation.

In every single-scan experiment the systematic noise $N(E_{kin})$ was introduced by switching-on the x-ray source when measuring in the range of $28.8-35.9$\,eV kinetic energies (shaded area in the bottom of Fig.\,2). A resulted single-scan spectrum in this energy region is shown by filled circles in Fig.\,2. In such way generated noise represents secondary electrons in the x-ray spectrum of the W(110) surface. We emphasize that this source of electrons is independent from the first one caused by the He\,II$\alpha$ radiation.

From this consideration we can rewrite the total photocurrent in the following form:
\begin{equation}
j(E_{kin})=I\cdot const \cdot DOS(E_{kin})+N(E_{kin}).
\end{equation}
For a series of measurements, the previous expression can be rewritten as
\begin{equation}
j_n=I_n\cdot\theta_{n}+N_n, n=1,2,\ldots,
\end{equation}
where $n$ is an iteration number. Observable (strictly measured) variables $j_n$ and $I_n$ correspond to $j(E_{kin})$ and $I$, while unobservable investigated variable $\theta_n$ and systematic noise $N_n$ correspond to $const \cdot DOS(E_{kin})$ and $N(E_{kin})$, respectively. Depending on even slight changes of the experimental conditions (like temperature drift, etc.) from one measurement to another, the investigated $const \cdot DOS(E_{kin})$ parameter can randomly vary within a certain distribution function around a fixed mean value. The goal of the whole measurement process is to probe this mean value. For each experiment we can consider unknown $\theta_n$ values to be randomly distributed:
\begin{equation}
\theta_n=\overline{\theta}+w_n, n=1,2,\ldots,
\end{equation}
where $\overline{\theta}$ is the investigated mean value and $w_n$ is a random zero-mean disturbance.

If $N_n$ from Eq. (4) represents a systematic noise with a non-zero mean value, conventionally used algorithms are failed to correctly estimate $\overline{\theta}$. To account for the non-zero mean contribution of SN to the estimated mean value of the investigated parameter~\cite{Granichin:2004}, in the present experiment randomly distributed $I_n$ were generated (see above).

Let us assume, the observable variable $I_n$ is randomly distributed around its known mean value $\overline{I}$. Note, there is no correlations between $I_n$ and $N_n$ that is of high importance for consistency of the proposed in the following algorithm. The latter is always true, if $I_n$ is selected to be random, while $N_n$ is unknown, but bounded deterministic function. Main issue of the suggested in Ref.~\cite{Granichin:2004} algorithm is to estimate mean value $\overline{\theta}$ of [$const \cdot DOS(E_{kin})$] measuring $j_n$ (experimental spectra) for randomly distributed $I_n$ (intensity of He\,II$\alpha$ radiation). Intuitively, the underlying mathematics can be understood on the basis of Equation (4), which can be partially differentiated with respect to variable $I_n$. In case $I_n$ and $N_n$ are independent from each other, a partial derivative of $j_n$ is equal to $\theta_n$.

Based on the above, the task of eliminating the systematic noise is reformulated to construct a sequence of estimations of average value of parameter of the linear regression model. It was shown~\cite{Granichin:2004} that in this case one can effectively use a SPSA type algorithm or randomized least-square method. 

In order to evaluate the experimental data we applied a modification of the SPSA algorithm which is described in Ref.~\cite{Granichin:2004} [see Fig.\,1(b) for the scheme of SPSA application]. Everywhere in the following we use $\hat{\theta}_n$ instead of $\theta_n$ to emphasize that the estimated, not real, values of the investigated variable are considered
\begin{equation}
\hat{\theta}_n=\hat{\theta}_{n-1}-\frac{\Delta_n}{\sigma^{2}_{n}n}(I_{n}\hat{\theta}_{n-1}-j_n), n=1,2,\ldots, \hat{\theta}_{0}=0,
\end{equation}
where $\Delta_n=I_n-\overline{I}$ and $\sigma^{2}_{n}$ is positive bounded dispersion of $\Delta_n$.

If the estimation sequence $\left\{\hat{\theta}_n\right\}$ leads to some particular value $\tilde{\theta}\geq0$ we can assert that this final approximate result equals to the real value of $\theta$ (for details see~\cite{Granichin:2004,Fedin:2007}). In the experiment, however, we can follow only finite number of measurements. Supposing that after some measurements the sequence of $\left\{\hat{\theta}_n\right\}$ becomes stable we can assume with high probability that this estimation value is in good agreement with the real one.

For our analysis we have used fifty PE spectra where randomly chosen SN was added. Figure\,3 shows the evolution of estimated values $\hat{\theta}_n$ which have to be proportional to the $DOS$ value. In order to prove the SPSA algorithm we chose twelve control points in PE spectra at particular kinetic energies (points are marked on kinetic energy axis in Fig.\,2). The intensities at these points in PE spectra during step-by-step application of the algorithm are plotted in Fig.\,3 with corresponding zoom for the last 10 steps (see inset). One can see that in spite of the overestimated large initial ``guess" values, the algorithm process is almost stabilized at around 20th iteration. From iteration 46th we have the most stable estimations consequence. Figure\,2 shows the result of the application of fifty steps of the SPSA algorithm (bottom spectrum). The evaluated spectrum is in good agreement with the one directly measured in the experiment, when the x-ray source was not activated (no SN noise). 

\textit{In conclusion}, we have demonstrated that the application of the SPSA algorithm is an effective way of SN attenuation in the linear regression case. As an example we applied this algorithm to filtering the systematic noise in the PE spectra. It was found that a set of 50 single-scan spectra is already enough to eliminate the systematic error. On the basis of these results we conclude that application of SPSA can be useful for analysis of a large variety of experimental data.

This work was supported by the DAAD under grant A/05/58613 and the Deutsche Forschungsgemeinschaft SFB 463, TP B4 and B16.

\newpage

\textbf{Figure captions:}
\newline
\newline
\textbf{Fig.\,1.} (a) Scheme of the PE experiment, where sample (1) is illuminated by two light sources, He\,II$\alpha$ and Al\,K$\alpha$, electrons are analyzed by photoelectron spectrometer (2) and detector (3) registers the signal $j_n$. (b) Layout of the present study using the SPSA algorithm for eliminating noises of unknown nature.
\newline
\newline
\textbf{Fig.\,2.} Experimental PE spectra of the valence band of W(110) obtained with He\,II$\alpha$ radiation without (open circles) and with (filled circles) systematic noise. Spectrum obtained after application of the SPSA algorithm to a series of 50 experimental single-scan spectra is shown by thin line. The shaded area in the bottom is systematic noise measured separately. Twelve control points used for demonstration of the convergence dynamic of the algorithm are matked with labels (1-12) on the kinetic energy axis. 
\newline
\newline
\textbf{Fig.\,3.} Convergence of the SPSA algorithm: The estimated values $[const\cdot DOS(E_{kin})]$ of the twelve control points marked in Fig.\,2. 

\clearpage
\begin{figure}[t]\center
\includegraphics[scale=0.3]{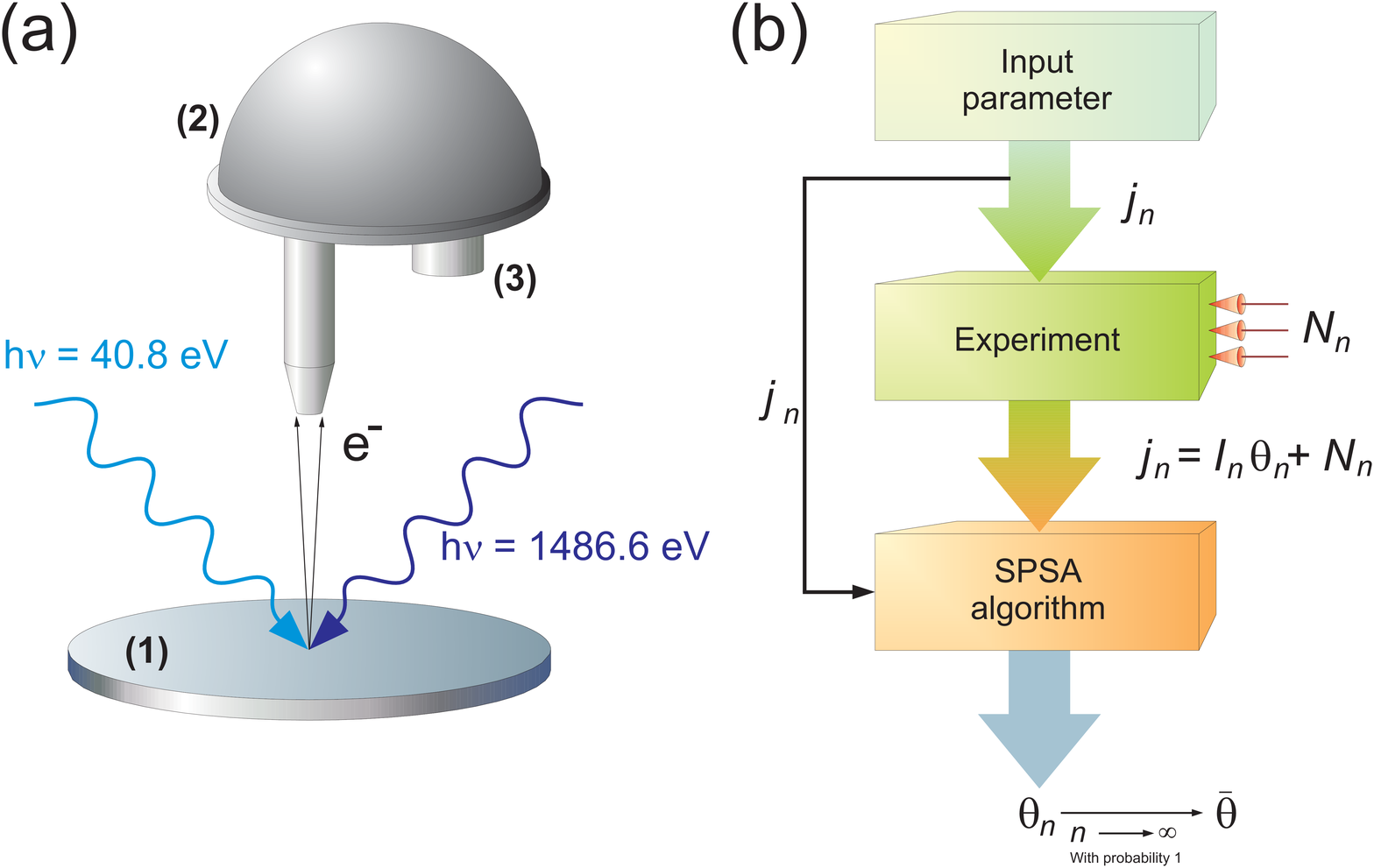}\\
\vspace{1cm}
\large \textbf{Fig.\,1.}
\end{figure}

\clearpage
\begin{figure}[t]\center
\includegraphics[scale=0.7]{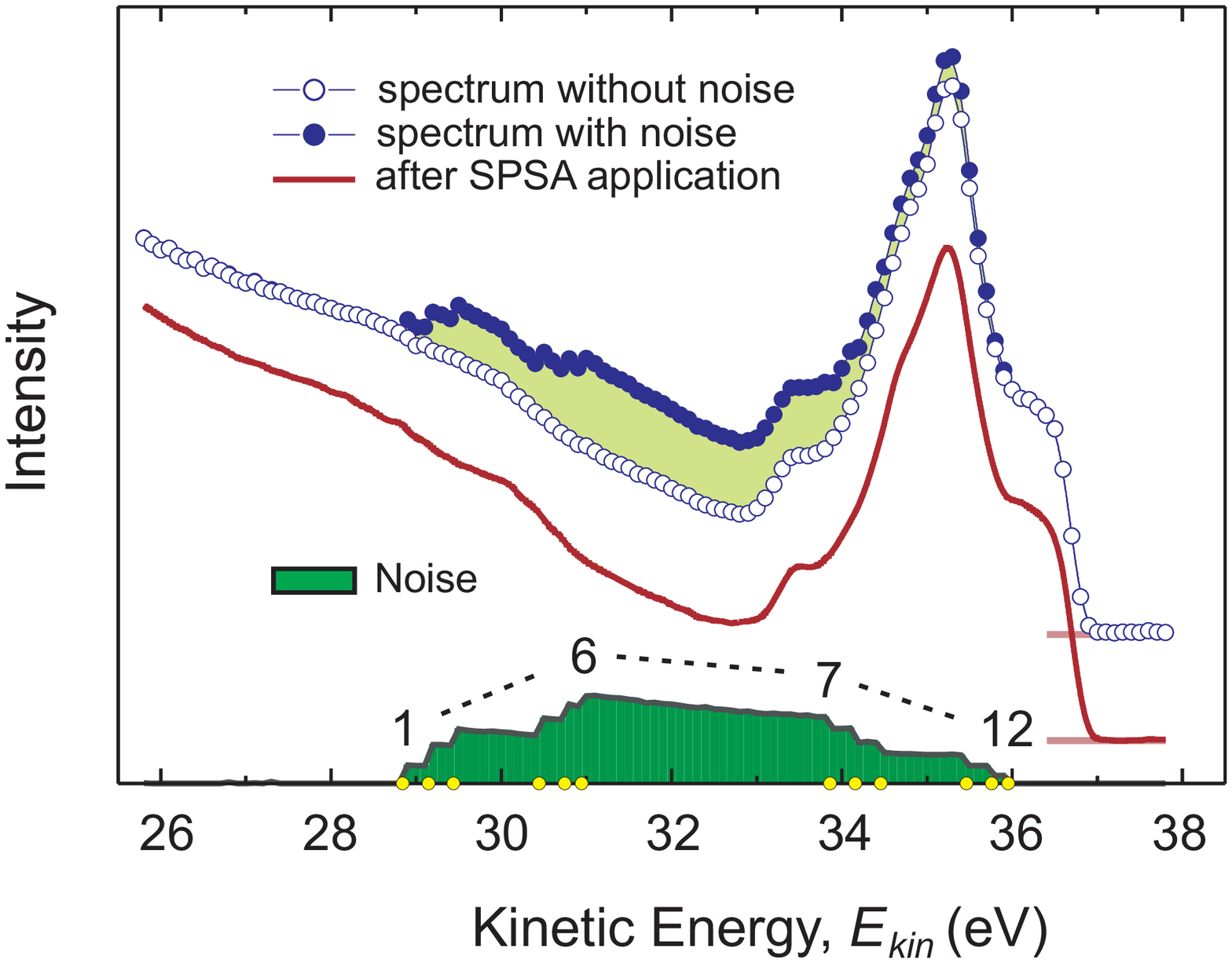}\\
\vspace{1cm}
\large \textbf{Fig.\,2.}
\end{figure}

\clearpage
\begin{figure}[t]\center
\includegraphics[scale=0.7]{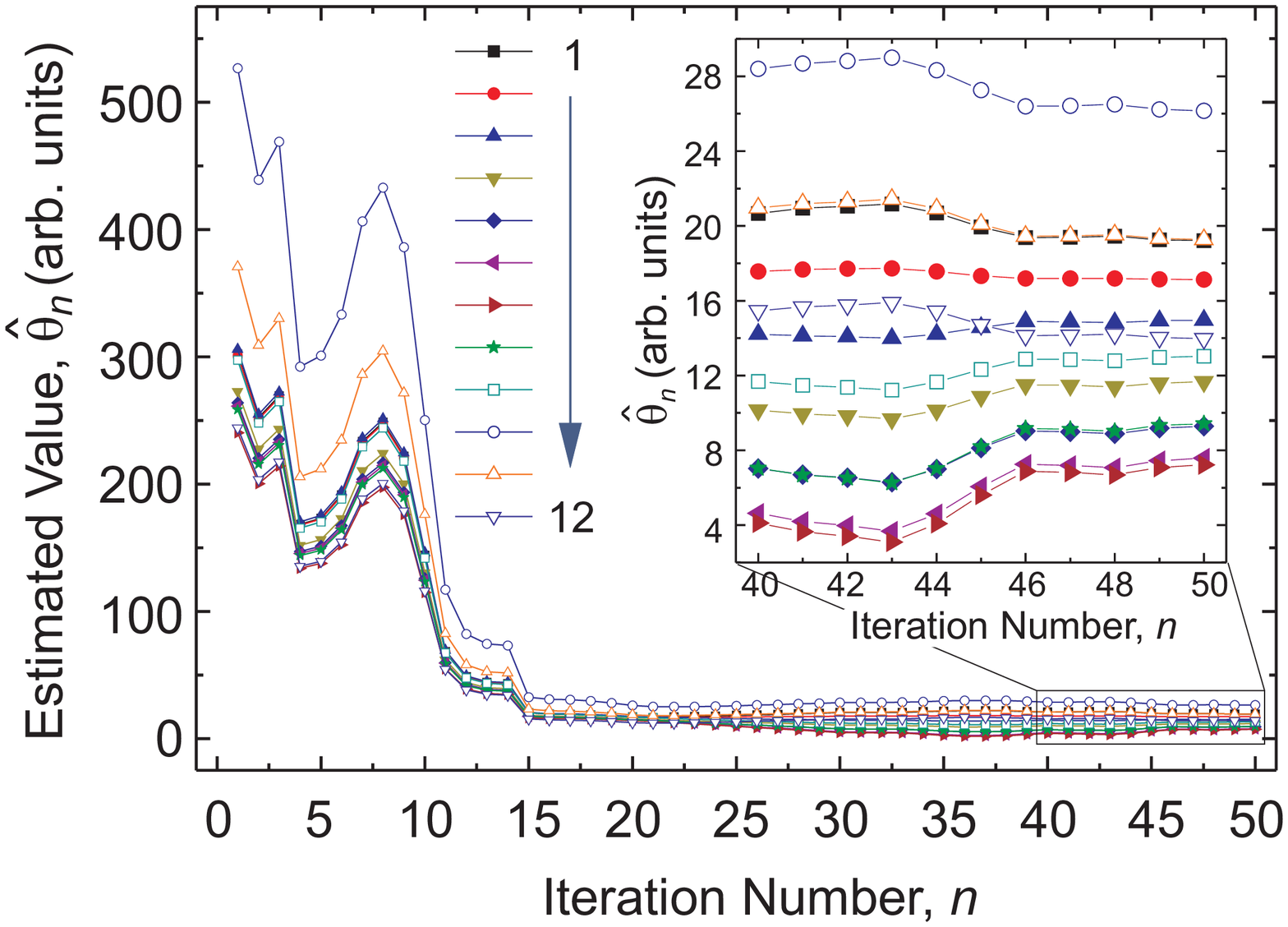}\\
\vspace{1cm}
\large \textbf{Fig.\,3.}
\end{figure}

\end{document}